\documentclass[showpacs,preprintnumbers,amsmath,amssymb,twocolumn]{revtex4}

\usepackage{amsmath}
\usepackage{graphicx}
\usepackage{epsf}
\usepackage{psfrag}
\usepackage{epsfig}
\usepackage{graphics}

\begin{document}
\newcommand{\e}{\mbox{e}}
\newcommand{\be}{\begin{equation}}
\newcommand{\ee}{\end{equation}}
\newcommand{\bq}{\begin{eqnarray}}
\newcommand{\eq}{\end{eqnarray}}
\newcommand{\intx}{\int^{1}_{0} dx}
\newcommand{\pbruto}{\hbox{$p \!\!\!{\slash}$}}
\newcommand{\eps}{\epsilon}
\newcommand{\qbruto}{\hbox{$q \!\!\!{\slash}$}}
\newcommand{\lbruto}{\hbox{$l \!\!\!{\slash}$}}
\newcommand{\kbruto}{\hbox{$k \!\!\!{\slash}$}}

\title{{\bf Regularization Independent Analysis of the Origin of Two Loop Contributions to N=1 Super Yang-Mills Beta Function}}

\date{\today}

\author{H. G. Fargnoli$^{(a)}$} \email[]{helvecio@fisica.ufmg.br}
\author{B. Hiller$^{(b)}$} \email[]{brigitte@teor.fis.uc.pt}
\author{A. P. Ba\^eta Scarpelli$^{(c)}$} \email[]{scarp@fisica.ufmg.br}
\author{Marcos Sampaio$^{(a)}$} \email []{msampaio@fisica.ufmg.br}
\author{M. C. Nemes$^{(a)}$}\email[]{mcnemes@fisica.ufmg.br}

\affiliation{(a) Federal University of Minas Gerais - Physics
Department - ICEx \\ P.O. BOX 702, 30.161-970, Belo Horizonte MG -
Brazil}
\affiliation{(b) Coimbra University - Faculty of Science and Technology - Physics Department - Center of Computational Physics\\
Rua Larga, P-3004-516 Coimbra - Portugal}
\affiliation{(c)Setor T\'ecnico-Cient\'ifico - Departamento de Pol\'icia Federal - Rua Hugo D'Antola, 95 - Lapa - S\~ao Paulo - Brazil}

\begin{abstract}

\noindent
We present a both ultraviolet and infrared regularization independent analysis in a symmetry preserving framework for the $N=1$ Super Yang-Mills beta function to two loop order. We show explicitly that off-shell infrared divergences as well as the overall two loop ultraviolet divergence cancel out whilst the beta function receives contributions of infrared modes. 

\end{abstract}


\maketitle

\section{Introduction}
%%%%%%%%%%%%%%%%%%%%%%%%%%%%%%%%%%%%%%%%%%%%%%%

In this contribution we aim at shedding some light, from the perturbative standpoint, on the origin of loop contributions to the beta function of $N=1$ Supersymmetric Yang-Mills (SYM) theories. We employ an automatically invariant, four dimensional (minimal) framework in which neither Ward identities need to be imposed as constraint equations nor modifications in the original physical Lagrangian are performed. Furthermore both finite and divergent parts of the Feynman amplitudes are displayed which is particularly useful to discuss the subtleties involved in the corrections of the beta function beyond one loop order in this model as we shall discuss later on.

The beta function of SYM theory has been a subject of controversy in both non-perturbative and perturbative calculations.  One of the debates has been named ``anomaly puzzle".  Because the $U(1)_R$ current is in the same multiplet as the trace of the energy momentum tensor, chiral anomaly and trace anomaly are equally subjected to the Adler-Bardeen theorem \cite{Adler}. As the beta function is proportional to the trace of the energy momentum tensor, one could assert on general grounds that contributions beyond one loop order were zero. On the other hand Novikov, Shifman, Vainshtein and Zakharov (NSVZ) were the first to obtain an exact expression for the beta function \cite{Novikov}. In addition, perturbative calculations using dimensional reduction \cite{Siegel}
revealed higher loop contributions to the beta function. Other regularization methods have obtained corrections beyond one loop as well \cite{Mas}, \cite{Pimenov}, \cite{Abdalla}. Arkani-Hamed and Murayama \cite{Murayama} have shed some light on the anomaly puzzle providing a simple and elegant explanation. They have shown that the vector multiplet does not have canonical kinetic terms after dilations (trace anomaly) which can only appear after an additional rescaling is performed. Hence the anomaly from this modified dilation  will no longer belong to the same multiplet as the $U(1)_R$ anomaly. As a result, the beta function corrections are not constrained by the Adler Bardeen theorem. However, some controversies remain \cite{Huang}.

The other controversy on this matter concerns the possible infrared mode contributions to the beta function which again affects both non-perturbative and perturbative calculations.  NSVZ \cite{Novikov} derived the exact beta function in a framework based on instanton analysis which borne out its infrared origin.  Contrariwise, Arkani-Hamed and Murayama \cite{Murayama} claim, within a purely Wilsonian framework freed from infrared subtleties, that the exact beta function depends only on ultraviolet properties of the theory. This discussion also appears in a perturbative analysis. Within the supergraph approach to supersymmetric models, along with on-shell infrared divergences of Yang-Mills theory, additional off-shell infrared divergences appear \cite{Abbott} which must be {\it{consistently}} separated from ultraviolet ones before renormalization is effected. The unclear distinction between these two types of infinities is at the heart of this  debate. The answer to this matter passes through an unambiguous distinction between the infinities involved and the arbitrary scales which are byproducts of the subtractions.
In dimensional reduction \cite{Abbott},\cite{Grisaru} the two loop correction to the beta function appears from a local evanescent operator typical of the method. Such operator is absent should the calculation stay in the physical space-time dimension. Therefore Grisaru, Milewski and Zanon themselves conjectured that no divergence should occur beyond one loop. This is correct as we verify later in this contribution. However it does not mean that the two loop beta function vanishes but instead that the naive perturbative derivation of the beta function based on renormalization constants needs some reinterpretation and modification. Such modification appears to be related to scaling anomalies \cite{Kraus}. It is important to observe that in dimensional regularizations the treatment of amplitudes which are simultaneously infrared and ultraviolet divergent is involved because they can be mixed.
Mas, Perez-Victoria and Seijas have evaluated the two loop contribution within Differential Renormalization \cite{Mas} which does not recourse to a dimensional extension on the space-time dimension. In this framework no infinities appear by construction. They claim that the beta function two loop correction depends on infrared divergences yet the latter is seen as playing a passive role. We believe that a calculation in  which 1) no dimensional continuation on space-time is performed; 2) infrared and ultraviolet divergences decouple clearly; 3) one distinguishes the corresponding scales without the need of  introducing extra parameters to ensure their independence; 4)  gauge and supersymmetry invariance are maintained  at every loop order; 5) only one renormalization scale appears as divergences are subtracted and 6) one can keep track of both finite and infinite parts of the amplitude, can shed more light on the perturbative analysis of this problem.

Implicit Regularization is a good candidate for such a task. It has been proved to be invariant to all orders if well known surface terms are set to zero which is closely related to momentum routing invariance in the Feynman diagrams \cite{IR},\cite{Mota},\cite{Edson},\cite{Adriano}. Ultraviolet infinities need not be evaluated and remain as  a systematic set of basic divergent integrals which depend upon a unique renormalization scale $\lambda$. Their derivatives with respect to that scale are also basic divergent integrals. Off-shell infrared infinities can be likewise defined save that the basic divergent integrals depend on an infrared arbitrary scale which is independent of $\lambda$. Moreover, it was shown that Implicit Regularization complies with locality, unitarity and Lorentz invariance because the Bogoliubov's recursion relations which locally executes the forest formula to subtract  general subdivergences can be implemented in such a scheme.

In the present contribution we calculate the two loop beta function of $N=1$ Super Yang-Mills theory in an ultraviolet and off-shell infrared regularization independent way in the framework of Implicit Regularization. We verified in our calculation at two loop order the conjecture by Grisaru, Milewski and Zanon \cite{Grisaru} that after subtraction of subdivergences no ultraviolet infinities occur beyond one loop in N=1 Super Yang-Mills when space-time dimension is not modified. More importantly we conclude that the two first terms in the loop expansion of the beta function receive contributions related to on-shell infrared divergences.

\section{Pure N=1 Supersymmetric Yang-Mills}

Consider the standard pure $N=1$ supersymmetric Yang-Mills classical action \cite{Mas}, \cite{Grisaru} 
\begin{equation*}
S_{SYM} = \frac{1}{g^{2}}\int d^4 x d^2 \theta\ \mbox{Tr}W^2 .
\end{equation*}
We work with  the supersymmetric background field method where gauge invariance is kept manifest.  The one loop contributions depicted by the diagrams of figure 1 read, after performing supersymmetric algebra,
\begin{eqnarray*}
D_A &=& -\frac{3}{2}C_{A}\int d^{4}p\ d^{4}\theta\ \widetilde{\mathbf{W}}^{\alpha}(p)\widetilde{\mathbf{\Gamma}}_{\alpha}(-p) \nonumber \\ &\times& \int^{\Lambda_{uv}}\frac{d^4 k}{(2\pi)^4} \frac{1}{k^{2}(p+k)^{2}}\\
D_B &=& i\frac{\xi}{4}C_{A} \int d^{4}p\ d^{2}\theta\ \widetilde{\mathbf{W}}^{\alpha}(p)\widetilde{\mathbf{W}}_{\alpha}(-p)p^{2} \nonumber \\ &\times& \int^{\Lambda_{ir}}\frac{d^4 k}{(2\pi)^4}\frac{1}{k^{4}(k-p)^{2}}
\end{eqnarray*}
where $\xi$ is a gauge parameter and the tilde indicates the Fourier transform. The superscripts $\Lambda_{uv}, \Lambda_{ir}$ indicate that the integrals are implicitly regularized in the ultraviolet and infrared limits so to allow for algebraic manipulations with the integrand. Hereafter, for brevity, we adopt the notation $\int_{k}\equiv \int^{\Lambda_{uv}, \Lambda_{ir}}\frac{d^4 k}{(2\pi)^4}$.

\begin{figure}[ht]
\begin{center}
\includegraphics [height=1.9 cm]{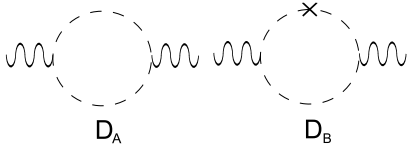}
\end{center}
\caption{One loop diagrams of the two-point function.}
\label{4pontos}
\end{figure}

By simple power counting we conclude that the first integral in $k$ is only ultraviolet divergent while the second one is only infrared divergent. In order to avoid as long as possible contact with infrared regularization schemes we leave the infrared integrals in their original form assuming an adequate regulator that allows for algebraic manipulations. In the appendix we specify two concrete infrared regularizations. We can write $D_A$ in terms of the classical action after integrating over half of the supercoordinates, to obtain
\begin{eqnarray}
D_A &=& \frac{3}{4}C_{A}i\int d^{4}p\ d^{2}\theta\ \widetilde{\mathbf{W}}^{\alpha}(p)\widetilde{\mathbf{W}}_{\alpha}(-p) I(p) \label{1loopA}\\
D_B &=& i\frac{\xi}{4}C_{A} \int d^{4}p\ d^{2}\theta\ \widetilde{\mathbf{W}}^{\alpha}(p)\widetilde{\mathbf{W}}_{\alpha}(-p)p^{2} U(p) \nonumber\\
&& \label{1loopB}
\end{eqnarray}
where we have defined for later use
\begin{equation*}
U(p)\equiv \int_{k}\frac{1}{k^{4}(k-p)^{2}}\quad , \quad I(p) \equiv \int_{k}\frac{1}{k^2(p-k)^2}.
\end{equation*}

The ultraviolet divergent integral $I(p)$ can be written in the convenient form using Implicit Regularization (see for instance \cite{Edson})
\begin{eqnarray}
I(p) &=& I_{log}(\lambda^{2})-b\ln\left(-\frac{p^{2}}{\lambda^{2}}\right)+2b\\
&\equiv& I_{log}(\lambda^{2}) + I^{fin}\left(p^2, \lambda^2\right)
\end{eqnarray}
where $b\equiv\frac{i}{(4\pi)^2}$, $\lambda$ is a nonvanishing arbitrary parameter which plays the role of renormalization group scale and
\begin{equation}\label{ilog}
I_{log}(\lambda^{2})\equiv \int_{k}\frac{1}{(k^2-\lambda^2)^2}
\end{equation}
is a one loop basic divergent integral \footnote{At $n$th loop order the basic divergent integrals are written as powers up to $n$ of $I_{log}(\lambda^2)$ and $I_{log}^{(n)}(\lambda^2)= \int d^4k/(2 \pi)^4 \ln^{n-1}[(k^2-\lambda^2)/(-\lambda^2)]/(k^2-\lambda^2)^2$ \cite{IR},\cite{Edson},\cite{Adriano}. Similar expressions appear for quadratic and linearly divergent integrals.}.

\begin{figure}[ht]
\begin{center}
\includegraphics [height=3.8 cm]{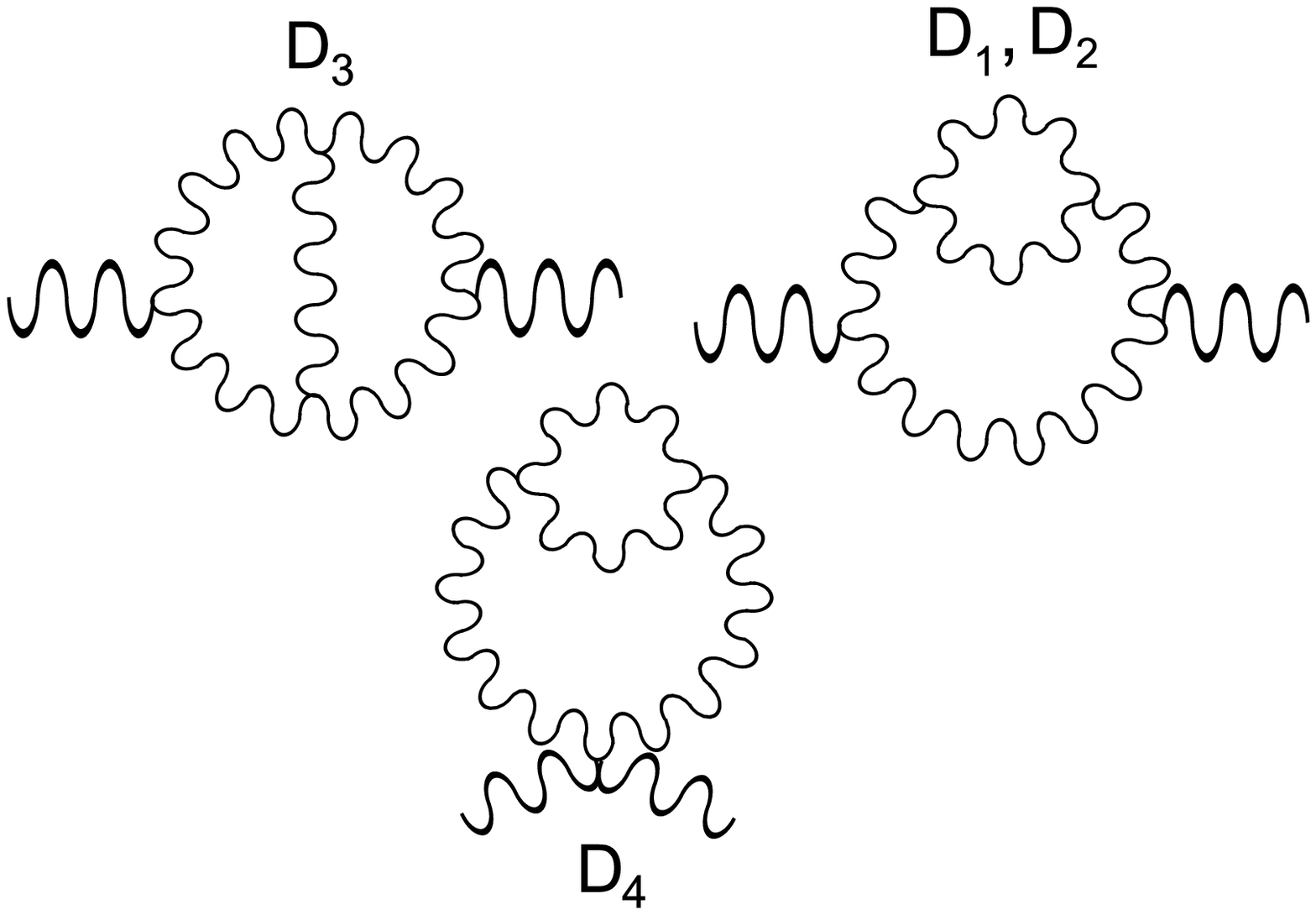}
\end{center}
\caption{Two loop diagrams of the two-point function.}
\label{2pontos}
\end{figure}

The two loop contributions depicted in figure 2 can be written, omitting a common factor $\frac{-3g^{2}C_{A}^{2}}{2}$, as
\begin{eqnarray*}
D_1 &=& \int d^{4}p\ d^{4}\theta\  \widetilde{\overline{\mathbf{W}}}^{\stackrel{.}{\alpha}}(p,\theta)\widetilde{\mathbf{W}}^{\alpha}(-p,\theta)  \sigma^{\mu}_{\alpha\stackrel{.}{\alpha}}\nonumber\\
&&\times \int_{k,q} \frac{(p-k)_{\mu}}{k^{4}(p-k)^{2}q^{2}(k-q)^{2}}\\
D_2 &=& \frac{1}{2}\int d^{4}\theta\ d^{4}p\ \widetilde{\mathbf{\Gamma}}^{\mu}(-p,\theta)\widetilde{\mathbf{\Gamma}}^{\nu}(p,\theta)\nonumber\\
&&\times \left[p_{\mu}p_{\nu}\int_{k,q}\frac{1}{q^{4}(p+q)^{2}}\frac{1}{k^{2}(k+q)^{2}}\right.\\
&&\left.+4\int_{k,q}\frac{(p+q)_{\nu}q_{\mu}}{q^{4}(p+q)^{2}}\frac{1}{k^{2}(q+k)^{2}}\right]\label{ddois}\\
D_3 &=& \frac{1}{2}\int d^{4}\theta\ d^{4}p\ \widetilde{\mathbf{\Gamma}}^{\mu}(-p,\theta)\widetilde{\mathbf{\Gamma}}^{\nu}(p,\theta)\nonumber\\
&&\times \left[\int_{k,q}\frac{(p+q)_{\mu}(p+k)_{\nu}}{q^{2}k^{2}(p+q)^{2}(p+k)^{2}(k-q)^{2}}\right.\\
&&\left.-\int_{k,q}\frac{(p+q)_{\mu}(p+k)_{\nu}}{(p+q+k)^{2}(p+k)^{2}q^{2}k^{2}(p+q)^{2}}\right]\\
D_4 &=& -\frac{g_{\mu\nu}}{2}\int d^{4}\theta\ d^{4}p\ \widetilde{\mathbf{\Gamma}}^{\mu}(-p,\theta)\widetilde{\mathbf{\Gamma}}^{\nu}(p,\theta)\nonumber\\
&&\times \int_{k,q}\frac{1}{(q+k)^{4}q^{2}k^{2}}.
\end{eqnarray*}

$D_1$, $D_2$ and $D_4$ are both ultraviolet and infrared power counting divergent while $D_3$ is only ultraviolet divergent. In addition $D_1$, $D_2$ and $D_4$ contain ultraviolet subdivergences which must be subtracted.

We start by evaluating $D_1$
\begin{eqnarray*}
D_1 &=& \int d^{4}p\ d^{4}\theta\  \widetilde{\overline{\mathbf{W}}}^{\stackrel{.}{\alpha}}(p,\theta)\widetilde{\mathbf{W}}^{\alpha}(-p,\theta) \sigma^{\mu}_{\alpha\stackrel{.}{\alpha}}\\
&&\times \int_{k} \frac{(p-k)_{\mu}}{k^{4}(p-k)^{2}}I(k).
\end{eqnarray*}
The ultraviolet subdivergence is removed minimally by subtracting $I_{log}(\lambda^2)$ from $I(k)$. We define $D'_1$ by substituting $I(k) \rightarrow I^{fin}\left(p^2, \lambda^2\right)$ in the expression above. Then, after using the identity $\mbox{\boldmath $\nabla$}^{\alpha}\widetilde{\mathbf{W}}_{\alpha}=\mbox{\boldmath$\overline{\nabla}$}^{\stackrel{.}{\alpha}}
\widetilde{\overline{\mathbf{W}}}_{\stackrel{.}{\alpha}}$, $D'_1$ becomes
\begin{eqnarray*}
D'_1&=&\int d^{4}p\ d^{2}\theta\  \widetilde{\mathbf{W}}_{\alpha}(p,\theta)\widetilde{\mathbf{W}}^{\alpha}(-p,\theta) p^{2}\nonumber\\
&&\times \left[-\int_{k} \frac{I^{fin}\left(k^2, \lambda^2\right)}{k^{4}(p-k)^{2}} +\frac{b}{p^{2}}I^{fin}\left(p^2, \lambda^2\right)\right].
\end{eqnarray*}

It is straightforward to see that $D_2$ can be cast as
\begin{eqnarray*}
&&D_2 = \frac{1}{2}\int d^{4}\theta\ d^{4}p\ \widetilde{\mathbf{\Gamma}}^{\mu}(-p,\theta)\widetilde{\mathbf{\Gamma}}^{\nu}(p,\theta)\nonumber\\
&&\times\left[p_{\mu}p_{\nu}\int_{q}\frac{I(q)}{q^{4}(p+q)^{2}} +4p_{\nu}\int_{q}\frac{q_{\mu}I(q)}{q^{4}(p+q)^{2}}\right.\nonumber\\
&&\left.+4\int_{q}\frac{q_{\nu}q_{\mu}}{q^{4}(p+q)^{2}}I(q)\right]
\end{eqnarray*}
and again $D'_2$ is obtained from $D_2$ substituting $I(q)$ by $I^{fin}\left(q^2, \lambda^2\right)$.

Consider now the amplitude $D_3$ which contain no subdivergences. After some simple algebra we find
\begin{eqnarray*}
D'_3 &\equiv& D_3 = \frac{1}{2}\int d^{4}\theta\ d^{4}p\ \widetilde{\mathbf{\Gamma}}^{\mu}(-p,\theta)\widetilde{\mathbf{\Gamma}}^{\nu}(p,\theta)\\
&&\times \left[-\frac{p_{\mu}p_{\nu}}{2}J(p)+2I_{\mu\nu}\right]
\end{eqnarray*}
where
\begin{equation*}
I_{\mu\nu} = \int_k\int_q
\frac{k_{\mu}q_{\nu}}{k^{2}q^{2}\left(k-p\right)^{2}\left(k-q\right)^{2}\left(q-p\right)^{2}}
\end{equation*}
and the finite integral \cite{Rosner}
\begin{eqnarray*}
J(p) &=& \int_k\int_q
\frac{1}{k^{2}q^{2}\left(k-p\right)^{2}\left(k-q\right)^{2}\left(q-p\right)^{2}}\nonumber \\
&=& - \frac{6 \zeta (3)}{(4 \pi)^4 p^2}.
\end{eqnarray*}

As for $D_4$ after changing  variables $q \rightarrow q' = k + q$ we separate it in  ultraviolet and infrared divergent pieces by multiplying and dividing by $(p+q)^2$. It results
\begin{eqnarray*}
D_4 &=& -\frac{g_{\mu\nu}}{2}\int d^{4}\theta\ d^{4}p\ \widetilde{\mathbf{\Gamma}}^{\mu}(-p,\theta)\widetilde{\mathbf{\Gamma}}^{\nu}(p,\theta)\\
&&\times \left[p^{2}\int_{q} \frac{I(q)}{q^{4}(p+q)^{2}}+\int_{q} \frac{I(q)}{q^{2}(p+q)^{2}}\right.\\
&&\left.+2p_\alpha\int_{q} \frac{q^{\alpha}I(q)}{q^{4}(p+q)^{2}}\right].
\end{eqnarray*}
$D'_4$ is obtained as before replacing $I(q)$ by $I^{fin}\left(q^2, \lambda^2\right)$.

The following integrals are contained in the expressions for $D'_i$'s. We present the results \cite{Edson} up to surface terms which we systematically set to zero to implement gauge (and momentum route) invariance as discussed in the introduction \cite{Edson}:
\begin{eqnarray*}
&&\int_{q}\frac{1}{q^{2}(q+p)^{2}}\ln\left(-\frac{q^{2}}{\lambda^{2}}\right)=I_{log}^{(2)}(\lambda^{2})+b\ln\left(-\frac{p^{2}}{\lambda^{2}}\right)\\
&&-\frac{b}{2}\ln^{2}\left(-\frac{p^{2}}{\lambda^{2}}\right),\\
&&\int_{q}\frac{q^{\alpha}}{q^{4}(q+p)^{2}} = -b\frac{p^{\alpha}}{p^{2}},\\
&&\int_{q}\frac{q^{\alpha}}{q^{4}(q+p)^{2}}\ln\left(-\frac{q^{2}}{\lambda^{2}}\right) = -b\frac{p^{\alpha}}{p^{2}}\ln\left(-\frac{p^{2}}{\lambda^{2}}\right),\\
&&\int_{q}\frac{q_{\mu}q_{\nu}}{q^{4}(q+p)^{2}} = \frac{g_{\mu\nu}}{4}\left[I_{log}(\lambda^{2})-b\ln\left(-\frac{p^{2}}{\lambda^{2}}\right)+2b\right]\\
&&+\frac{b}{2}\frac{p_\mu p_\nu}{p^{2}},\\
&&\int_{q}\frac{q_{\mu}q_{\nu}}{q^{4}(q+p)^{2}}\ln\left(-\frac{q^{2}}{\lambda^{2}}\right) = \frac{g_{\mu\nu}}{4}\left[I^{(2)}_{log}(\lambda^{2})\right.\\
&&\left.+\frac{1}{2}I_{log}(\lambda^{2})-\frac{b}{2}\ln^{2}\left(-\frac{p^{2}}{\lambda^{2}}\right)+\frac{b}{2}\ln\left(-\frac{p^{2}}{\lambda^{2}}\right)+\frac{b}{2}\right]\\
&&+\frac{p_\mu p_\nu}{p^{2}}\left[\frac{b}{4}+\frac{b}{2}\ln\left(-\frac{p^{2}}{\lambda^{2}}\right)\right],\\
&&I_{\mu\nu} = \frac{g_{\mu\nu}}{4}\left[b I_{log}(\lambda^{2})-b^{2}\ln\left(-\frac{p^{2}}{\lambda^{2}}\right)-\frac{p^{2}}{3}J(p)\right.\nonumber\\
&&\left.+\frac{11}{3}b^{2}-b^{2}\frac{\pi^{2}}{9}\right]+\frac{p_\mu p_\nu}{p^{2}}\left[\frac{p^{2}}{3}J(p)-\frac{1}{6}b^{2}+b^{2}\frac{\pi^{2}}{36}\right],
\end{eqnarray*}
where $I_{log}^{(2)}(\lambda^{2})=\int_{k}\frac{1}{(k^2-\lambda^2)^2}\ln\left(-\frac{k^{2}-\lambda^{2}}{\lambda^{2}}\right)$ and $I_{log}(\lambda^{2})$ is defined in (\ref{ilog}). 

The partial sum 
\begin{eqnarray}
&&D'_2+D'_3+D'_4 = \frac{1}{2}\int d^{4}\theta\ d^{4}p\ \widetilde{\mathbf{\Gamma}}^{\mu}(-p,\theta)\widetilde{\mathbf{\Gamma}}^{\nu}(p,\theta)\nonumber\\
&&\times \left(p_{\mu}p_{\nu}-p^{2}g_{\mu\nu}\right)\left\{\int_{q}\frac{I^{fin}\left(q^2, \lambda^2\right)}{q^{4}(p+q)^{2}}+\frac{b^{2}\zeta(3)}{p^{2}}\right.\nonumber\\
&&-\left.\frac{2b}{p^{2}}\left(-b\ln\left(-\frac{p^{2}}{\lambda^{2}}\right)+\frac{8}{3}b-\frac{b\pi^{2}}{36}\right)\right\}
\end{eqnarray}
is transverse as it should because of gauge invariance which is manifest in the background field method, showing that our calculation is consistent (invariant). Therefore it can be written in terms of the classical action by integrating over half of the supercoordinates to yield
\begin{eqnarray}
&&D'_2+D'_3+D'_4 = \frac{3}{4}\int d^{2}\theta\ d^{4}p\ \widetilde{\mathbf{W}}^{\alpha}(-p,\theta)\nonumber\\
&&\times \widetilde{\mathbf{W}}_{\alpha}(p,\theta) p^{2}\left\{\int_{q}\frac{I^{fin}\left(q^2, \lambda^2\right)}{q^{4}(p+q)^{2}}+\frac{b^{2}\zeta(3)}{p^{2}}\right.\nonumber\\
&&\left.-\frac{2b}{p^{2}}\left(-b\ln\left(-\frac{p^{2}}{\lambda^{2}}\right)+\frac{8}{3}b-\frac{b\pi^{2}}{36}\right)\right\}.
\end{eqnarray}
Now we can add all the contributions to the two loop two-point function as
\begin{eqnarray}
&&\sum_{i=1}^{4}D'_i = \int d^{2}\theta\ d^{4}p\ \widetilde{\mathbf{W}}^{\alpha}(-p,\theta)\widetilde{\mathbf{W}}_{\alpha}(p,\theta)\nonumber\\
&&-\frac{3g^2C_{A}^{2}}{2}\left[-\frac{1}{4}p^{2}\int_{k} \frac{I^{fin}\left(k^2, \lambda^2\right)}{k^{4}(p-k)^{2}}\right.\nonumber\\
&&\left. +\frac{b^2}{2}\ln\left(-\frac{p^{2}}{\lambda^{2}}\right)+\frac{3}{4}b^{2}\zeta(3)-2b^{2}+\frac{b^{2}\pi^{2}}{24}\right] \label{total}.
\end{eqnarray}

Notice that the ultraviolet divergences at two loop order cancel out magically in the sum $D'_1 +D'_2+D'_3+D'_4$ as predicted by Grisaru, Milewski and Zanon \cite{Grisaru}. As we have said in the introduction, this does not mean that the two loop beta function coefficient vanishes. In order to evaluate the beta function one can use the renormalization group equation. Adding the tree, one loop and two loop contributions given by (\ref{1loopA}), (\ref{1loopB}) and (\ref{total}) allows us to write the (ultraviolet) renormalized two-point Green function as
\begin{eqnarray}\label{final}
&&G_{ren}^{(2)}(p^{2})= \frac{1}{2g^{2}}\nonumber\\
&&+\frac{C_{A}}{4}i\left[-3b\ln\left(-\frac{p^{2}}{\lambda^{2}}\right) + 6b + \xi p^{2}U(p) \right]\nonumber\\
&&-\frac{3g^2C^{2}_{A}}{8}\left[-2bp^{2}U(p)+bp^2 U^{(2)}(p)\right.\nonumber\\
&&\left.+2b^{2}\ln\left(-\frac{p^{2}}{\lambda^{2}}\right)-8b^{2}+3b^{2}\zeta(3)+\frac{b^{2}\pi^{2}}{6}\right]
\end{eqnarray}
where the first term is the tree level contribution,  $U(p)$ was defined in the one loop calculation and $U^{(2)} (p) $ stands for
\begin{equation}
U^{(2)}(p) \equiv \int_{k}\frac{1}{k^{4}(k-p)^{2}}\ln\left(-\frac{k^{2}}{\lambda^{2}}\right).
\end{equation}
Notice that $U^{(2)}(p)$ is off-shell infrared divergent and it contains both an off-shell infrared ($\tilde{\lambda}$) and ultraviolet ($\lambda$) scales (see the appendix). They are obviously independent by construction.  Moreover one can easily check that
$$
\lambda \frac{\partial}{\partial \lambda} \Big( G_{ren}^{(2)} (p^2) \Big)_{\tilde{R}}=
\Bigg( \lambda \frac{\partial}{\partial \lambda} G_{ren}^{(2)} (p^2)\Bigg)_{\tilde{R}}
$$
which guarantees the independence of the scales $\lambda$ and $\widetilde{\lambda}$. Here $\tilde{R}$ is the operation that subtracts off-shell infrared infinities. We leave the off-shell infrared divergences (implicitly regularized) represented by $U(p)$ and $U^{(2)}(p)$ in an integral form to show explicitly that they play no role in the evaluation of the beta function.

$G_{ren}^{(2)} (p^2)$ obeys the renormalization group equation \footnote{If the infrared and ultraviolet were not independent a term proportional to $\widetilde{\lambda}\frac{\partial}{\partial \widetilde{\lambda}}$ should be added in (\ref{renorm}), $\widetilde{\lambda}$ being the infrared scale.}:
\begin{equation}\label{renorm}
\left(\lambda\frac{\partial}{\partial \lambda}+\beta(g)\frac{\partial}{\partial g}+\gamma_{\xi}\frac{\partial}{\partial \xi}\right)G_{ren}^{(2)}(p^{2})\bigg|_{\xi=0} = 0 .
\end{equation}
Using $\gamma_{\xi}=-\frac{3C_A}{(4\pi)^{2}}g^{2}+\mathcal{O}(g^4)$ \cite{Mas} and (\ref{final}) we can solve the above equation for $\beta(g)$. Writing $\beta(g) = b_1 g^3 + b_2 g^5 +\mathcal{O}(g^7)$ we find
\begin{eqnarray}
b_1 &=& -\frac{3}{4}\frac{C_A}{(4\pi)^2}\label{betaum}\\
b_2 &=& -\frac{3C_{A}^{2}}{8}\left[bp^2\lambda^{2}\frac{\partial}{\partial\lambda^2}U^{(2)}(p)-2b^2+bp^2 U(p)\right].\nonumber\\
&& \label{betadois}
\end{eqnarray}
Note that the second term on the right hand side of (\ref{betadois}) comes from the term proportional to $\ln(-p^2/\lambda^2)$ of the last line of (\ref{final}). The terms $U^{(2)}(p)$ and $U(p)$ are separately off-shell infrared divergent. However $\lambda^2 \frac{\partial}{\partial\lambda^2}U^{(2)}(p) = - U(p)$ independently of the infrared regulator and thus
\begin{equation}\label{segunda}
b_2 = -\frac{3}{4}\frac{C^{2}_{A}}{(4\pi)^4}.
\end{equation}
It is important to note that such cancelation does not depend on the infrared regularization and since just $U^{(2)}(p)$ and $U(p)$ are (off-shell) infrared divergent the two first universal coefficients of the beta function are independent of off-shell infrared infinities. Indeed in the appendix we explicitly evaluate the integrals $U(p)$ and $U^{(2)}(p)$ in two different infrared regularizations showing such cancelation. The above beta function coefficients agree with the literature (see for instance \cite{Mas}).

\section{Concluding Remarks}

We used, we believe for the first time, a four dimensional invariant framework in which gauge and supersymmetry invariance are {\it{ automatically}} maintained, and both finite and infinite parts of the amplitudes are made explicit throughout the calculation, without any modification to the original effective action (i.e. with the simplest Feynman rules), to study the loop contributions to the beta function of $N=1$ SYM theory. In the Feynman gauge, along with on-shell infrared divergences, additional off-shell divergences appear as expected \cite{Abbott}; they  must be carefully dealt with, in a way that they do not get mixed with ultraviolet ones. This is achieved through the identification of distinct scales in connection with the $R$ and/or $\tilde{R}$ operations.

As conjectured in \cite{Grisaru} for four dimensional invariant methods, we verified within our framework that after subdivergences are subtracted, no overall ultraviolet divergences survive at two loop order. However the beta function does not vanish as calculated with the finite part of the amplitude which obeys a regular renormalization group equation.

As it becomes clear from our equation (\ref{betadois}), off-shell infrared divergences cancel out automatically in a (off-shell infrared) regularization independent fashion (that is to say, no $\tilde{R}$ operation is necessary) and hence there is no contribution whatsoever stemming from them. We have explicitly verified within two off-shell infrared regularization frameworks the cancelation of the first and third term in (\ref{betadois}) using in one case a dual version of Implicit Regularization in coordinate space and in the other case a simple fictitious mass in the propagators (see appendix). However the term in the middle, proportional to $b^2$, stems from the derivative of a non local on-shell infrared divergence proportional to $\ln(-p^2/\lambda^2)$ which multiplies the $\widetilde{\mathbf{W}}(p,\theta)$ fields and is integrated over the whole $p$ four dimensional space. Therefore it becomes apparent that (on-shell) infrared modes contribute to the two loop beta function. This is somewhat not surprising as it may happen in other supersymmetric theories \cite{West}.

It is clear that a naive perturbative derivation of the beta function from renormalization constants must be modified taking into account the presence of the rescaling anomalies discussed in \cite{Kraus}. Our approach reveals itself as very adequate for such analysis.
\vspace{0.2cm}

\textbf{Acknowledgements.} B. Hiller acknowledges valuable discussions with Alexander A. Osipov. M. Sampaio thanks Manuel Perez-Victoria for enlightening suggestions. H. G. Fargnoli, A. P. Ba\^eta Scarpelli, M. Sampaio and M. C. Nemes acknowledge financial support by CNPq.

\section*{APPENDIX}

In \cite{Brigitte3} the integrals $U(p)$ and $U^{(2)}(p)$ were evaluated using a dual version of Implicit Regularization that deals with infrared divergences. Using their results:
\begin{eqnarray*}
U&=&\frac{1}{p^{2}}\left(\widetilde{I}_{log}(\widetilde{\lambda}^{-2})+b\ln\left(-\frac{p^{2}}{\overline{\widetilde{\lambda}}^{2}}\right)+2b\right)\\
U^{(2)}&=& \frac{1}{p^{2}}\ln\left(\frac{\overline{\widetilde{\lambda}}^{2}}{\lambda^{2}}\right)\left(\widetilde{I}_{log}(\widetilde{\lambda}^{-2})+b\ln\left(-\frac{p^{2}}{\overline{\widetilde{\lambda}}^{2}}\right)\right.\\
&+&2b\Bigg)+\frac{b}{p^2} \left[\frac{1}{2} \ln^2 \left(
\frac{-p^2}{\overline{\widetilde{\lambda}}^{2}}\right) + \ln \left(
\frac{-p^2}{\overline{\widetilde{\lambda}}^2}\right)\right]\\
&&-\frac{1}{p^{2}}\widetilde{I}_{log}^{(2)}(\widetilde{\lambda}^{-2})
\end{eqnarray*}
where $\widetilde{\lambda}$ is an infrared scale independent of the ultraviolet one, $\overline{\widetilde{\lambda}}\equiv \left(4 /\e^{2\gamma}\right)\widetilde{\lambda}$ with $\gamma$ being the Euler-Mascheroni constant. Here $\widetilde{I}_{log}(\widetilde{\lambda}^{-2})$ and $\widetilde{I}_{log}^{(2)}(\widetilde{\lambda}^{-2})$ are infrared basic divergent integrals (analogous to the ultraviolet ones). Substituting these results in (\ref{betadois}) we see the cancelation of the infrared divergent pieces and the nonlocal terms and we get the correct value of $b_2$ (\ref{segunda}).

Alternatively one can introduce an infrared mass regulator $\mu$ in the propagators finding:
\begin{eqnarray*}
U &=& \int_{k}\frac{1}{(k^2-\mu^2)^2[(p-k)^{2}-\mu^2]}\\
&=& \frac{b}{p^2}\left[\ln\left(-\frac{p^2}{\widetilde{\lambda}^2}\right)+\ln\left(\frac{\widetilde{\lambda}^2}{\mu^2}\right)\right] + \mathcal{O}(\mu^2)\\
U^{(2)} &=&  \int_{k}\frac{1}{(k^2-\mu^2)^2[(p-k)^{2}-\mu^2]}\ln\left(-\frac{k^2-\mu^2}{\lambda^{2}}\right)\nonumber\\
&=&\frac{b}{p^2}\left\{\ln\left(-\frac{p^2}{\widetilde{\lambda}^2}\right)+\ln\left(\frac{\widetilde{\lambda}^2}{\mu^2}\right)\right.\\
&+&\frac{1}{2}\ln^{2}\left(-\frac{p^2}{\widetilde{\lambda}^2}\right)-\frac{1}{2}\ln^{2}\left(\frac{\widetilde{\lambda}^2}{\mu^2}\right)\\
&+&\left.\left[\ln\left(-\frac{p^2}{\widetilde{\lambda}^2}\right)+\ln\left(\frac{\widetilde{\lambda}^2}{\mu^2}\right)\right]\ln\left(\frac{\widetilde{\lambda}^2}{\lambda^2}\right)\right\}+ \mathcal{O}(\mu^2)
\end{eqnarray*}
Again substituting these results in (\ref{betadois}) we get the correct result of $b_2$.


\begin{thebibliography}{99}

\bibitem{Adler} S. L. Adler and W. A. Bardeen, \textit{Phys. Rev.} \textbf{182} (1969) 1517.

\bibitem{Novikov} V. Novikov, M. Shifman, A. Vainstein, V.
Zakharov, \textit{Phys. Lett.} \textbf{B 166} (1985) 329, \textit{Nucl. Phys.} \textbf{B 229} (1983) 381.

\bibitem{Siegel} W. Siegel, \textit{Phys. Lett.} \textbf{B 84}(1979) 193; O. V. Tarasov, V. A. Vladimirov, \textit{Phys. Lett.} \textbf{B 96} (1980) 94; M. T. Grisaru, M. Rocek and W. Siegel, \textit{Phys. Rev. Lett.} \textbf{45} (1980) 1063; W. Caswell and  D. Zanon, \textit{Phys. Lett.} \textbf{B 100}(1980) 152.

\bibitem{Mas} J. Mas, M. Perez-Victoria, C. Seijas, \textit{JHEP} \textbf{0203} (2002) 049.

\bibitem{Pimenov} A. B. Pimenov, E. S. Chevtsova and K. V. Stepanyantz, \textit{Phys. Lett.} \textbf{B 686} (2010) 293.

\bibitem{Abdalla} E. Abdalla and R. S. Jasinschi, \textit{Nucl. Phys.} \textbf{B 286} (1987) 42.

\bibitem{Murayama} N. A-Hamed and H. Murayama, \textit{JHEP} \textbf{0006} (2000) 030.

\bibitem{Huang} X. Huang and L. Parker, \textit{Eur. Phys. J. } \textbf{C} \textbf{71} (2011) 1570.

\bibitem{Abbott} L. F. Abbott, M. T. Grisaru, D. Zanon, \textit{Nuc. Phys.} \textbf{B 244} (1984) 454; M. T. Grisaru, D. Zanon, \textit{Nuc. Phys.} \textbf{B 252} (1985) 357.

\bibitem{Grisaru} M. T. Grisaru, B. Milewski and D. Zanon, \textit{Phys. Lett.} \textbf{B 155} (1985) 357.

\bibitem{Kraus} E. Kraus, \textit{Phys. Rev.} \textbf{D 65} (2002) 105003.

\bibitem{IR} C. R. Pontes, A. P. Ba\^eta Scarpelli, Marcos Sampaio, J. L.
Acebal, M. C. Nemes, \textit{Eur. Phys. J. C}\textbf{53}(2008)121; C. R. Pontes, A. P. Ba\^eta Scarpelli, Marcos Sampaio and M. C. Nemes \textit{J. Phys. G} \textbf{34}, (2007) 2215; L. C. T. Brito, H. Fargnoli, A. P. Ba\^eta
Scarpelli, Marcos Sampaio, M. C. Nemes \textit{Phys. Lett. B} \textbf{673}, (2009) 220; A. P. Ba\^{e}ta Scarpelli, M. Sampaio and M. C. Nemes, \textit{{Phys. Rev. \textbf{D}}} \textbf{63} (2001) 046004; A. P. Ba\^{e}ta Scarpelli, M. Sampaio, B. Hiller and M. C. Nemes, \textit{{Phys. Rev. \textbf{D}}} \textbf{64} (2001) 046013; M. Sampaio, A. P. Ba\^eta Scarpelli, B. Hiller, A. Brizola, M. C. Nemes and S. Gobira, \textit{{Phys. Rev. \textbf{D}}} \textbf{65} (2002) 125023; S. R. Gobira and M. C. Nemes, \textit{ Int. J. Theor. Phys.} \textbf{42} (2003) 2765;D. Carneiro, A. P. Ba\^eta Scarpelli, M. Sampaio and M. C. Nemes, \textit{JHEP} \textbf{12} (2003) 044; M. Sampaio, A. P. Ba\^eta Scarpelli, J. E. Ottoni, M. C.
Nemes, \textit{Int. J. Theor. Phys.} \textbf{45} (2006) 436; L. A. M. Souza, Marcos Sampaio, M. C. Nemes, \textit{Phys. Lett.} \textbf{B 632} (2006) 717; J. E. Ottoni, A. P. Ba\^eta Scarpelli, Marcos Sampaio, M. C. Nemes, \textit{Phys. Lett.} \textbf{B 642} (2006) 253; E. W. Dias, B. Hiller, A. L. Mota, M. C. Nemes, Marcos Sampaio, A. A. Osipov, \textit{Mod. Phys. Lett.} \textbf{A 21} (2006) 339; Brigitte Hiller, A. L. Mota, M. C. Nemes, A. A. Osipov, Marcos Sampaio, \textit{Nucl. Phys.} \textbf{A 769} (2006) 53.

\bibitem{Mota} O. A. Battistel, A. L. Mota, M. C. Nemes \textit{{Mod.
Phys. Lett. \textbf{A}}} \textbf{13} (1998) 1597.

\bibitem{Edson} E. W. Dias, A. P. Ba\^eta Scarpelli, L. C. T. Brito, Marcos Sampaio, M. C. Nemes,
\textit{Eur. Phys. J. } \textbf{C} \textbf{55} (2008) 667.

\bibitem{Adriano} A. Cherchiglia, Marcos Sampaio and M. C. Nemes, accepted for publication in \textit{Int. J. Mod. Phys.} \textbf{A}, 1008.1377v3 [hep-th].

\bibitem{Rosner} J. L. Rosner, \textit{Ann. Phys.} \textbf{44}
(1967) 11.

\bibitem{Brigitte3} H. G. Fargnoli, A. P. Ba\^eta Scarpelli, L. C. T. Brito, B. Hiller, Marcos Sampaio, M. C. Nemes and A. A. Osipov, \textit{Mod. Phys. Lett.} \textbf{A 26} (2011) 289.

\bibitem{Shifman} M. A. Shifman and A. I. Vainshtein, \textit{Nucl. Phys.} \textbf{B 277} (1986) 456.

\bibitem{West} P. West, \textit{Introduction to Supersymmetry and Supergravity}, World Scientific Publishing Company, Second Edition, 1990.

\end{thebibliography}
\end{document}